\documentclass[pra,twocolumn,floatfix,a4paper,superscriptaddress]{revtex4-2}

\usepackage{bm,color,amsmath,txfonts}
\usepackage{graphicx}
\usepackage{siunitx}
\usepackage{subfigure}
\usepackage{verbatim}
\usepackage{dcolumn}% Align table columns on decimal point
\usepackage{bm}% bold math
\usepackage{epsf}
\usepackage{xcolor}
\usepackage{hyperref}
\usepackage{hhline}
\usepackage{braket}
\usepackage{float}
\usepackage{enumerate}
\usepackage{bbm}
\usepackage{lipsum}
\usepackage{appendix}
\usepackage{mathtools}

\hypersetup{colorlinks=true, citecolor=blue, urlcolor=blue, linkcolor=blue}
\begin{document}

%\preprint{APS/123-QED}

\title{Entanglement distribution among distinct mechanical nodes in a quantum network}% Force line breaks with \\

\author{Zhi-Yuan Fan}
\thanks{fanzhiyuan@sxnu.edu.cn}%
 \affiliation{School of Physics and Electronic Engineering, Shanxi Normal University, Taiyuan, Shanxi 030031, China}%Lines break automatically or can be forced with \\
 
\author{Liu-Yong Cheng}%
\affiliation{School of Physics and Electronic Engineering, Shanxi Normal University, Taiyuan, Shanxi 030031, China}
\affiliation{Institute of Quantum Science and Technology, Yanbian University, Yanji 133002, China}%

\date{\today}% It is always \today, today,

\begin{abstract}
We propose two schemes to achieve remote entanglement distribution between two mechanical nodes with a significant frequency mismatch, based on optomechanical systems. The first scheme utilizes the physical mechanism to redistribute the quantum entanglement initially established in a dispersively-coupled optomechanical system with a megahertz mechanical resonance to a distant optomechanical system which embodies the triple-resonant interaction induced by the scattering of gigahertz mechanical phonon. We also provide a fast optical pulse protocol to realize the long-distance entanglement distribution from the optomechanical system supporting the gigahertz mechanical mode to the megahertz mechanical mode included in a distant optomechanical system. Specifically, these two schemes respectively demonstrate the megahertz-to-gigahertz and gigahertz-to-megahertz entanglement distribution in the quantum network of optical photons and phonons. This work may facilitate the application of various mechanical systems in hybrid quantum network-based quantum technologies and  fundamental physical research.
\end{abstract}

\maketitle

\section{Introduction}
The emergence of hybrid quantum systems highlights the necessity to utilize diverse quantum systems with distinct advantages and complementary functionalities to cooperatively perform the quantum tasks \cite{Ladd2010,Georgescu2014,Bennett2000,Degen2017,Kimble2008,Wehner2018,Azuma2023}. 
As one indispensable building block, the mechanical systems, which support long lifetime, fabrication flexibility in size and resonance frequency (range from kilohertz to gigahertz), may find broad applications in modern quantum technologies \cite{Kurizki2015,Clerk2020}. Therefore, it is of significant importance to possess the alibity to manipulate and measure the physical state of different mechanical systems. The study of interactions between light and matter opened up a burgeoning field to achieve this, namely, the cavity optomechanics \cite{Aspelmeyer2014}. Along with the promising progress in micro-nano fabrication, optical manipulation and precision measurement, the nonlinear interaction between photons and mechanical phonons has been demonstrated on a variety of optomechanical physical platforms in recent decades, which enables optical control and readout of the mechanical motions in these hybrid systems. 
In the quantum regime, the cavity optomechanics has  unique application in realizing various kinds of mechanical quantum states on the macroscopic scale \cite{Palomaki2013,Riedinger2016,Hong2017,Marinkovic2018,Fiaschi2021,Wollman2015,Pirkkalainen2015,Lecocq2015,Chan2011,Teufel2011,Rossi2018,Clark2017,Park2009,Schliesser2008,Delic2020,Li2018A,Vanner2021,Vanner2021B,Qiu2020}. With regard to the mechanical entangled states, there has been many schemes in advances of cavity optomechanics \cite{Li2022,Jiao2024,Wang2016,Chen2017,Chakraborty2018,Hu2020,Clarke2020,Joshi2012,Wang2013,Liao2014,Yang2015,Pirandola2006,Hartmann2008}. Specifically, the experimental demonstration of macroscopic entanglement between a couple of mechanical systems with gigahertz \cite{Riedinger2018} or megahertz \cite{Ockeloen-Korppi2018,Lepinay2021,Kotler2021} resonances have been achieved due to their vital important applications in quantum field. The high frequency mechanical systems with gigahertz resonance have an obvious superiority on the robustness against thermal effect and play crucial roles in quantum fundamental test \cite{Hong2017,Marinkovic2018}, quantum teleportation \cite{Fiaschi2021}, low noise microwave to optics conversion \cite{Forsch2019}, and quantum computational devices interface \cite{Mirhosseini2020}, etc. The megahertz-resonant mechanical systems, with long coherent lifetime, also find unique applications in quantum information storing \cite{Fiore2011}, precision measurement \cite{Schliesser2009,Hertzberg2010,Ockeloen-Korppi2016}, nonreciprocal device fabrication \cite{Shen2016,Ruesink2016,Bernier2017,Peterson2017}, and mechanical phonon lasing \cite{Grudinin2010,Jing2014,Zhang2018}. 
%non-Gaussian quantum state preparation \cite{Li2018A,Qiu2020},
%quantum squeezed state \cite{Wollman2015,Pirkkalainen2015,Lecocq2015} and ground state \cite{Teufel2011,Clark2017,Rossi2018,Park2009,Schliesser2008} preparation,
In view of this, it is natural to consider combining such two distinct systems and further realizing entanglement or even remote entanglement of two mechanical systems with a significant frequency mismatch, which leads to entanglement distribution between different mechanical nodes in a quantum network. However, up to date, the mechanical entanglement between distinct mechanical nodes \cite{Tan2013,Jiao2022} or over a long distance \cite{Li2015,Bai2021} has been much less explored, which  restricts the wide application of distinct mechanical systems in hybrid quantum network constuction and related quantum technologies. 
%. As a result, achieving distant quantum entanglement between two different kind of mechanical nodes 

Here, we show that the obstacle can be eliminated and  provide two theoretical proposals to achieve the macroscopic entanglement between distinct mechanical nodes. The first scheme mainly employs a hybrid model including two optomechanical systems, which embody mechanical resonances with a significant frequency mismatch, and respectively support dispersive and triple-resonant interactions. %In principle, by applying a {\it strong} red-detuned drive to the optomechanical system with low frequency (megahertz) mechanical mode, the quantum entanglement among optical cavity mode and  mechanical mode can be activated via the nonlinear dispersive coupling.
In principle, under the pumping of a {\it strong} red-detuned drive, the quantum entanglement between the optical cavity and  megahertz mechanical mode can be activated in the  optomechanical system with nonlinear dispersive coupling. Via the mediation of propagating optical photons, such quantum correlation can be further distributed to the gigahertz mechanical node exploiting
the Brillouin scattering (BLS)-induced optomechanical state transfer process in the cascaded optomechanical system, thus realizing distant entanglement distribution from megahertz to gigahertz mechanical nodes in the hybrid quantum network. 
Besides, how to distribute the entanglement from the optomechanical system which includes the gigahertz mechanical node to the remote mechanical node in megahertz range also merits great research interest. To reach this, we give an optical pulse-assisted scheme by successively activate the optomechanical Stokes and anti-Stokes processes. The former achieves quantum entanglement by preparing a two-mode squeezed vacuum state between the gigahertz mechanical node and the optical pulse. Then, the quantum state of the travelling temporal mode can be mapped onto the megahertz mechanical node through the state transfer activated by the anti-Stokes scattering in the remote optomechanical system, which establishes a nonlocal entanglement shared by distinct mechanical systems.

\begin{figure}[t]
	\centering
	\includegraphics[width=1\linewidth]{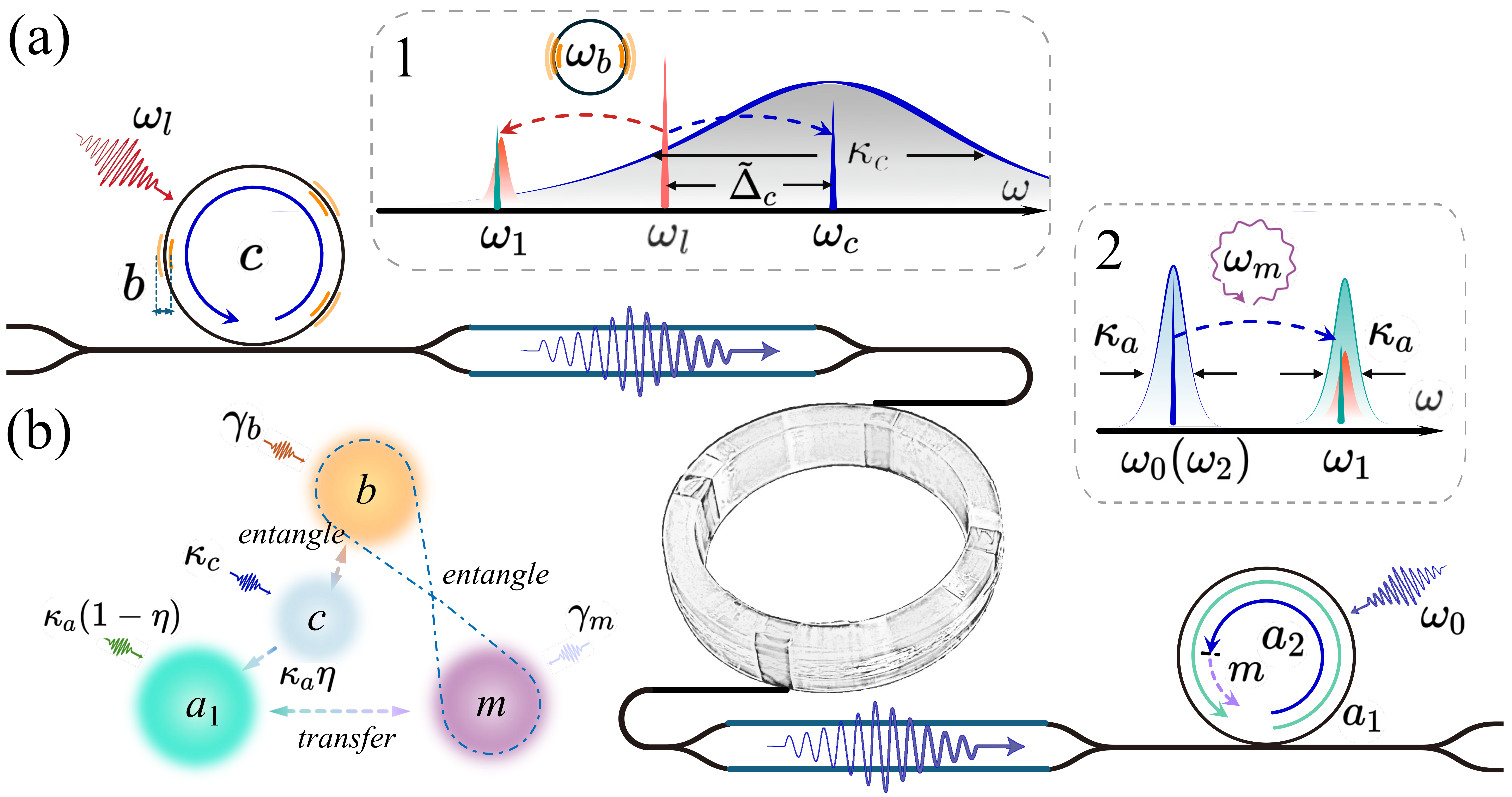}
	\caption{Schematic diagrams of the ($\mathrm{a}$)  hybrid system and ($\mathrm{b}$) physical mechanism to realize the remote megahertz-to-gigahertz mechanical entanglement distribution. Inset 1: Entanglement generation process. The down-conversion sideband ($\omega_l-\omega_b$) of the optical cavity mode ($c$) can couple to the WGM ($a_1$) via the mediation of optical waveguide. Inset 2: Entanglement distribution process. The optical WGM ($a_2$) is resonantly pumped by a laser to activate the beam-splitter type interaction.}
	\label{fig1}
\end{figure}

\section{Remote entanglement distribution from low- to high-frequency mechanical nodes}
\label{sec2}

\subsection{Theoretical Model}
As schematically shown in Fig.\ \ref{fig1}(a), the hybrid system under study is comprised of two cascaded optomechanical systems, which are connected by an optical waveguide and exhibit different optomechanical couplings. Specifically, the upstream system consists of a whisper gallery mode (WGM) and a mechanical mode with resonant frequency in the megahertz range. Considering the corresponding optomechanical configurations of microsphere \cite{Park2009,Fiore2011,Shen2016,Shen2022} and microdisk \cite{Schliesser2008,Schliesser2009,Weis2010,Zhang2018}, the optical cavity resonant frequency can experience a significant change induced by the mechanical oscillation, thus yielding the {\it dispersive} optomechanical interaction. Thanks to the strong localization on the electromagnetic field in these small-sized devices, the bare optomechanical coupling can get strengthened up to sub-kilohertz \cite{Aspelmeyer2014}. Via mediation of the optical waveguide in the telecom wavelength, the optical output field from the upstream optomechanical system can transmit to the downstream one, which supports two optical WGMs, a high-frequency mechanical mode in the gigahertz range, and the phonon BLS-induced {\it triple-resonant} interaction \cite{Vanner2021,Vanner2021B}. Actually, by resonantly driving any one cavity mode in this system, the BLS process can lead to a pronounced asymmetry between the Stokes and anti-Stokes sidebands. Hence, we can selectively activate the parametric-down conversion or beam-splitter (state transfer) interactions in this system. As a result, the Hamiltonian describing two separated systems in Fig. \ref{fig1}(a) reads (in units of $\hbar$)
%\begin{equation}
%	\begin{split}
%		H=&\ H_1+H_2,\\
%		H_1/\hbar =&\ \omega_c c^\dagger c+\omega_b b^\dagger b - g_c c^\dagger c(b+b^\dagger )+H_c^{\mathrm{dri}},\\
%		H_2/\hbar =&\ \omega_1 a_1^\dagger a_1 + \omega_2 a_2^\dagger a_2+\omega_m m^\dagger m \\ 
% &+ g(a_1^\dagger a_2m+a_1a_2^\dagger m^\dagger )+H_2^{\mathrm{dri}}, 
%	\end{split}
%\end{equation}
\begin{equation}
	\label{system}
	\begin{split}
		H=&\ H_f+H_\mathrm{int}+H_{\mathrm{dri}},\\
		H_f =&\ \omega_c c^\dagger c+\omega_b b^\dagger b+\omega_1^{} a_1^\dagger a_1^{} + \omega_2^{} a_2^\dagger a_2^{}+\omega_m m^\dagger m,\\
		H_\mathrm{int} =&\ -g_c^{} c^\dagger c(b+b^\dagger )+g(a_1^\dagger a_2^{} m+a_1^{} a_2^\dagger m^\dagger), \\
		H_{\mathrm{dri}} % &\ H_c^{\mathrm{dri}} +H_2^{\mathrm{dri}}\\
		=&\ i\epsilon(c^\dagger e^{-i\omega_l t}-ce^{i\omega_l t})+\ i\epsilon_2^{}(a_2^\dagger e^{-i\omega_0 t}-a_2^{} e^{i\omega_0 t}), 
	\end{split}
\end{equation}
where $\mathcal{O}_l=c$, $b$, ($\mathcal{O}_r=a_1$, $a_2$ and $m$) and $\mathcal{O}_l^\dagger$ ($\mathcal{O}_r^\dagger$) represent the mode operators of the optical cavity mode and low-frequency mechanical mode (two WGMs and high-frequency mechanical mode) embodied by the upstream (downstream) optomechanical system depicted in Fig.\ \ref{fig1}(a), respetively. $\omega_{\mathcal{O}_l(\mathcal{O}_r)}$ denote the corresponding resonant frequencies. The second term $H_\mathrm{int}$ involves dispersive and triple-resonant interactions between the photons and phonons in this hybrid model, with coupling strengths $g_c^{}$ and $g$. They are typically weak, but the efffective coupling strengths can be greatly enhanced by adopting the corresponding drives to excite the specific physical process. In the third term, $H_{\mathrm{dri}}'=i\hbar \epsilon(c^\dagger e^{-i\omega_l t}-ce^{i\omega_l t})$ describes a strong red-detuned pump applied on the optical cavity, where $\epsilon=\sqrt{\kappa_c P/\hbar\omega_l}$, jointly determined by the cavity mode dissipation $\kappa_c$, the dirve power $P$ and frequency $\omega_l$, introduces the coupling strength between the cavity and drive fields. $H_{\mathrm{dri}}''=i\hbar \epsilon_2^{}(a_2^\dagger e^{-i\omega_0 t}-a_2^{} e^{i\omega_0 t})$ characterizes the drive resonant to the optical WGM $a_2$, with cavity decay rate $\kappa_a$, the driving frequency $\omega_0$, and power $P_2$. 

We note that the Hamiltonian \eqref{system} is provided without considering the physical coupling between two remote optomechanical systems. Instead, the optical cavity modes in such two spatially separated microresonators can be effectively connected by an optical waveguide of telecom wavelength, thus facilitating the directional coupling that enables the propagation of the output field of upstream system $c_{\mathrm{out}}=\sqrt{\kappa_c} c-\tilde{c}_{\mathrm{in}}$ \cite{Gardiner1985} to the input of the downstream system but not vice versa. Here, $\tilde{c}_{\mathrm{in}}$ refers to both the optical drive $\epsilon$ of cavity mode $c$ and its input noise $c_{\mathrm{in}}$. In theory, the waveguide transmission procedure can be modeled by an equivalent beam-splitter \cite{Leonhardt1997,Li2010,Tan2021,Li2021}, such that the input field of optical WGM can be rewritten by
\begin{equation}
	\label{bs}
	\tilde{a}_{1,\mathrm{in} }=\sqrt{\eta}c_{\mathrm{out} }+\sqrt{1-\eta}a_{1,\mathrm{in}},
\end{equation}
where $\eta$ and $a_{1,\mathrm{in} }$ describes the coupling efficiency and input thermal noise. As a consequence, including the dissipation and input noise of each mode and working in the rotating frame with respet to $\hbar\omega_l (c^\dagger c + a_1^\dagger a_1^{})+\hbar\omega_0 a_2^\dagger a_2^{}+\hbar(\omega_l-\omega_0)m^\dagger m$, we can obtain the quantum Langevin equations (QLEs) of whole model
\begin{equation}
	\begin{split}
		\raisetag{12pt}
		\dot{c}=&-i\Delta_c c-\frac{\kappa_c}{2}c + ig_c^{} c (b+b^\dagger ) + \epsilon +\sqrt{\kappa_c} c_{\mathrm{in} },\\
		\dot{b}=&-i\omega_b b-\frac{\gamma_b}{2}b + ig_c^{} c^\dagger c+\sqrt{\gamma_b} b_{\mathrm{in}}, \\
		\dot{a}_1= &-i\Delta_1 a_1 -\frac{\kappa_a}{2}a_1 -ig a_2 m +\sqrt{\kappa_a}\tilde{a}_{1,\mathrm{in} } \\
		%=& -i\Delta_1 a_1 -\frac{\kappa_a}{2}a_1 -ig a_2 m +\sqrt{\eta\kappa_a} c_{\mathrm{out}}+\sqrt{(1-\eta)\kappa_a}a_{1,\mathrm{in} }^l,\\
		\dot{a}_2 = & -i\Delta_2 a_2 -\frac{\kappa_a}{2}a_2 -ig a_1 m^\dagger + \epsilon_2+\sqrt{\kappa_a} a_{2,\mathrm{in} },\\
		\dot{m} = &-i\Delta_m m -\frac{\gamma_m}{2} m -ig a_1 a_2^\dagger + \sqrt{\gamma_m}m_{\mathrm{in} },
	\end{split}
\end{equation}
where $\Delta_c=\omega_c-\omega_l$, $\Delta_1=\omega_1-\omega_l$, $\Delta_2=\omega_2-\omega_0$, $\Delta_m=\omega_m-(\omega_l-\omega_0)$, and $\gamma_b$ ($\gamma_m$) is the dissipation rate of the low (high)-frequency mechanical mode. Apart from $a_{1,\mathrm{in}}$ in Eq.\ \eqref{bs}, $c_{\mathrm{in}}$ and $a_{2,\mathrm{in}}$ also respresent the zero-mean input noises of corresponding optical WGMs in the hybrid system. In general, these operators obey the following correlation: $\langle k_{\mathrm{in}}(t) k_{\mathrm{in}}^\dagger(t') \rangle=[n_{k}(\omega_{k})+1]\delta(t-t') \approx \delta(t-t')$ ($k=c$, $a_1$ and $a_2$). As for the high-quality mechanical resonant modes, i.e. $Q_{k'}=\omega_{k'}/\gamma_{k'}\gg 1$ ($k'=b$ and $m$), we can further take the Markovian approximation \cite{VG2001} to describe the autocorrelation of noises: $\langle {k'_{\mathrm{in}}}(t) {k'_{\mathrm{in}}}^{\dagger}(t') \rangle =\langle {k'_{\mathrm{in}}}^{\dagger}(t){k'_{\mathrm{in}}}(t') \rangle + \delta(t-t')\approx [n_{k'}(\omega_{k'})+1]\delta(t-t')$, where $n_{k'}(\omega_{k'})=1/[\mathrm{exp}(\hbar\omega_{k'}/k_B T)-1]$ is the mean thermal excitation number of the specific mechanical mode at temperature $T$, with Boltzman constant $k_B$. 

\subsection{Linearization and Covariance Matrix}

In cavity optomechanics, a prerequisite to establish quantum entanglement is to enhance the nonlinear optomechanical interaction. To achieve this, we adopt to simultaneously pump the optical modes $c$ and $a_2$. Via the mediation of optical waveguide, the optical WGM $a_1$ can also attain a strong activation, which leads to $|\langle c\rangle|$, $|\langle a_1\rangle|$ and $|\langle a_2\rangle|\gg 1$. Therefore, when the system envolves into the steady state after a sufficient period, we can study the system dynamics by expanding each mode operator as the sum of its steady-state average and the quantum fluctuations: $\mathcal{O}_i=\langle {\mathcal{O}}_i\rangle +\delta \mathcal{O}_i$ ($i\in \{l,r\}$), which enables us to obtain the linearized QLEs about the quantum fluctuation terms of all modes in the hybird system. Following the definitions of quantum quadratures $\delta X_{\mathcal{O}_i}=(\delta \mathcal{O}_i+\delta \mathcal{O}_i^\dagger)/\sqrt{2}$, $\delta Y_{\mathcal{O}_i}=-i(\delta \mathcal{O}_i-\delta \mathcal{O}_i^\dagger)/\sqrt{2}$ %and their noise operators $X_{\mathcal{O}_i,\mathrm{in}}$, $Y_{\mathcal{O}_i,\mathrm{in}}$
, the linearized QLEs can be further rewritten as a time-dependent differential equation of the quadrature vector $u(t)=\{\delta X_c(t)$, $\delta Y_c(t)$, $\delta X_b(t)$, $\delta Y_b(t)$, $\delta X_{a_1}(t)$, $\delta Y_{a_1}(t)$, $\delta X_{a_2}(t)$, $\delta Y_{a_2}(t)$, $\delta X_m(t)$, $\delta Y_m(t)\}^{\mathrm{T}}$:
\begin{equation}
	\label{linearizedQLEs}
	\dot{u}(t)=Au(t)+n(t),
\end{equation}
where $n(t)$=$\{\sqrt{\kappa_c}X_{c,\mathrm{in}}(t), \sqrt{\kappa_c}Y_{c,\mathrm{in}}(t)$, $\sqrt{\gamma_b}X_{b,\mathrm{in}}(t)$, $\sqrt{\gamma_b}Y_{b,\mathrm{in}}(t)$, $\sqrt{\kappa_a}X_{\bar{a}_1,\mathrm{in}}(t),\sqrt{\kappa_a}Y_{\bar{a}_1,\mathrm{in}}(t),\sqrt{\kappa_a}X_{a_2,\mathrm{in}}(t),\sqrt{\kappa_a}Y_{a_2,\mathrm{in}}(t),\sqrt{\gamma_m}X_{m,\mathrm{in}}(t)$, $\sqrt{\gamma_m}Y_{m,\mathrm{in}}(t)\}^\mathrm{T}$ represents the vector of the input noises, among which $\bar{a}_{1,\mathrm{in}}=\sqrt{1-\eta}a_{1,\mathrm{in} }-\sqrt{\eta} c_{\mathrm{in}}$. More importantly, the drift matrix is provided as
\begin{widetext}
	\begin{equation}
		\label{drift}
		A=\begin{pmatrix}
			-\frac{\kappa_c}{2}  & \tilde{\Delta}_c & -2\mathrm{Im}G_c  & 0 & 0 & 0 & 0 & 0 & 0 & 0\\
			-\tilde{\Delta}_c & -\frac{\kappa_c}{2} & 2\mathrm{Re}G_c & 0 & 0 & 0 & 0 & 0 & 0 & 0\\
			0 & 0 & -\frac{\gamma_b}{2} & \omega_b & 0 & 0 & 0 & 0 & 0 & 0\\
			2\mathrm{Re}G_c & 2\mathrm{Im}G_c & -\omega_b & -\frac{\gamma_b}{2} & 0 & 0 & 0 & 0 & 0 & 0\\
			\sqrt{\kappa_a\kappa_c\eta} & 0 & 0 & 0 & -\frac{\kappa_a}{2} & \Delta_1 & \mathrm{Im}G_m & \mathrm{Re}G_m & \mathrm{Im}G_2 & \mathrm{Re}G_2\\
			0 & \sqrt{\kappa_a\kappa_c\eta} & 0 & 0 & -\Delta_1 & -\frac{\kappa_a}{2} & -\mathrm{Re}G_m & \mathrm{Im}G_m & -\mathrm{Re}G_2 & \mathrm{Im}G_2\\
			0 & 0 & 0 & 0 & -\mathrm{Im}G_m & \mathrm{Re}G_m & -\frac{\kappa_a}{2} & \Delta_2 & \mathrm{Im}G_1 & -\mathrm{Re}G_1\\
			0 & 0 & 0 & 0 & -\mathrm{Re}G_m & -\mathrm{Im}G_m & -\Delta_2 & -\frac{\kappa_a}{2} & -\mathrm{Re}G_1 & -\mathrm{Im}G_1\\
			0 & 0 & 0 & 0 & -\mathrm{Im}G_2 & \mathrm{Re}G_2 & \mathrm{Im}G_1 & -\mathrm{Re}G_1 & -\frac{\gamma_m}{2}  & \Delta_m\\
			0 & 0 & 0 & 0 & -\mathrm{Re}G_2 & -\mathrm{Im}G_2 & -\mathrm{Re}G_1 & -\mathrm{Im}G_1 & -\Delta_m & -\frac{\gamma_m}{2}
		\end{pmatrix},
	\end{equation}
\end{widetext}
where $G_c=g_c^{}\langle c\rangle$, $G_1=g\langle a_1\rangle$, $G_2=g\langle a_2\rangle$ and $G_m=g\langle m\rangle$ are the effective coupling strengths established by the strong pumps in this scheme, and $\langle c\rangle$, $\langle a_1\rangle$, $\langle a_2\rangle$ and $\langle m\rangle$ denote the corresponding steady-state excitations, which can be directly derived from the expanded equations for steady-state average:
\begin{equation}
	\label{stdystate}
	\begin{split}
		0=&-i\tilde{\Delta}_c\langle c\rangle -\frac{\kappa_c}{2}\langle c\rangle + \epsilon,\\ 
		0=&-i\Delta_1 \langle a_1\rangle -\frac{\kappa_a}{2}\langle a_1\rangle -ig\langle a_2\rangle \langle m\rangle + \sqrt{\kappa_a\kappa_c\eta}\langle c\rangle-\sqrt{\kappa_a}\langle c_{\mathrm{in}}\rangle\\
		=&-i\Delta_1 \langle a_1\rangle -\frac{\kappa_a}{2}\langle a_1\rangle -ig\langle a_2\rangle \langle m\rangle + \sqrt{\kappa_a\kappa_c\eta}\langle c\rangle-\sqrt{\kappa_a/\kappa_c }\epsilon,\\
		0=&-i\Delta_2\langle a_2\rangle -\frac{\kappa_a}{2}\langle a_2\rangle -ig\langle a_1\rangle \langle m\rangle^\ast+ \epsilon_2,\\
		0=&-i\Delta_1\langle m\rangle -\frac{\gamma_m}{2}\langle m\rangle -ig\langle a_1\rangle \langle a_2\rangle^\ast.
	\end{split}
\end{equation}
Besides, in Eqs.\ \eqref{drift} and \eqref{stdystate}, $\tilde{\Delta}_c=\Delta_c-g_c^{}\langle c\rangle (\langle b\rangle+\langle b\rangle^\ast)$ is the effective detuning of the WGM $a_1$ with respect to the optical drive at $\omega_l$, which includes the optical resonant frequency change under the disperisve  optomechanical interaction.
%, with steady state $\langle b\rangle =i g_c|\langle c\rangle|^2/(\gamma_b/2+i\omega_b)$.

Considering the gaussian nature of all input noises in this system, it can be anticipated that the quantum dynamics governed by the linearized QLEs  \eqref{linearizedQLEs} will finally evolve into the gaussian state, which can be characterized by a $10\times10$ covariance matrix (CM), with each element defined by $V_{ij}=\langle u_i(t)u_j(t')+u_j(t')u_i(t)\rangle/2$ ($i,j\in\{1,2,...,10\}$). Specifically, the steady-state CM can be obtained by solving the so-called Lyapunov equation \cite{Vitali2007}
\begin{equation}
	A V+V A^{\mathrm{T}}+D=0.
\end{equation}
It is obvious that except drift matrix $A$, the steady-state CM $V$ also depends on the diffusion matrix $D$ which denotes the corresponding CM of different noises in vector $n(t)$. With each entry defined by $D_{ij}\delta(t-t')=\langle n_i(t)n_j(t')+n_j(t')n_i(t)\rangle/2$, it takes the form of 
\begin{equation}
	D=\begin{pmatrix}
		D_c & 0 & D_{ca_1}\\
		0 & D_b & 0\\
		D_{ca_1} & 0 & D_{a_1}
	\end{pmatrix}\oplus \begin{pmatrix}
		D_{a_2} & 0\\
		0 & D_m
	\end{pmatrix},
\end{equation}
where $D_c=\kappa_c(n_c+1/2)I_2$, $D_b=\gamma_b(n_b+1/2)I_2$, $D_{a_1}=\kappa_a[(1-\eta)(n_c+1/2)+\eta(n_{a_1}+1/2)]I_2$, $D_{a_2}=\kappa_a(n_{a_2}+1/2)I_2$, $D_m=\gamma_m(n_m+1/2)I_2$, including $2\times 2$ identity matrix $I_2$. Moreover, the off-diagonal part $D_{ca_1}=-\sqrt{\kappa_a\kappa_c\eta}(n_c+1/2)I_2$ features the correlation between input noises of $c_{\mathrm{in}}$ and $\bar{a}_{1,\mathrm{in}}=\sqrt{1-\eta} a_{1,\mathrm{in} }-\sqrt{\eta} c_{\mathrm{in}}$.

\subsection{Results and Discussion}
\begin{figure}[t]
	\centering
	\includegraphics[width=1\linewidth]{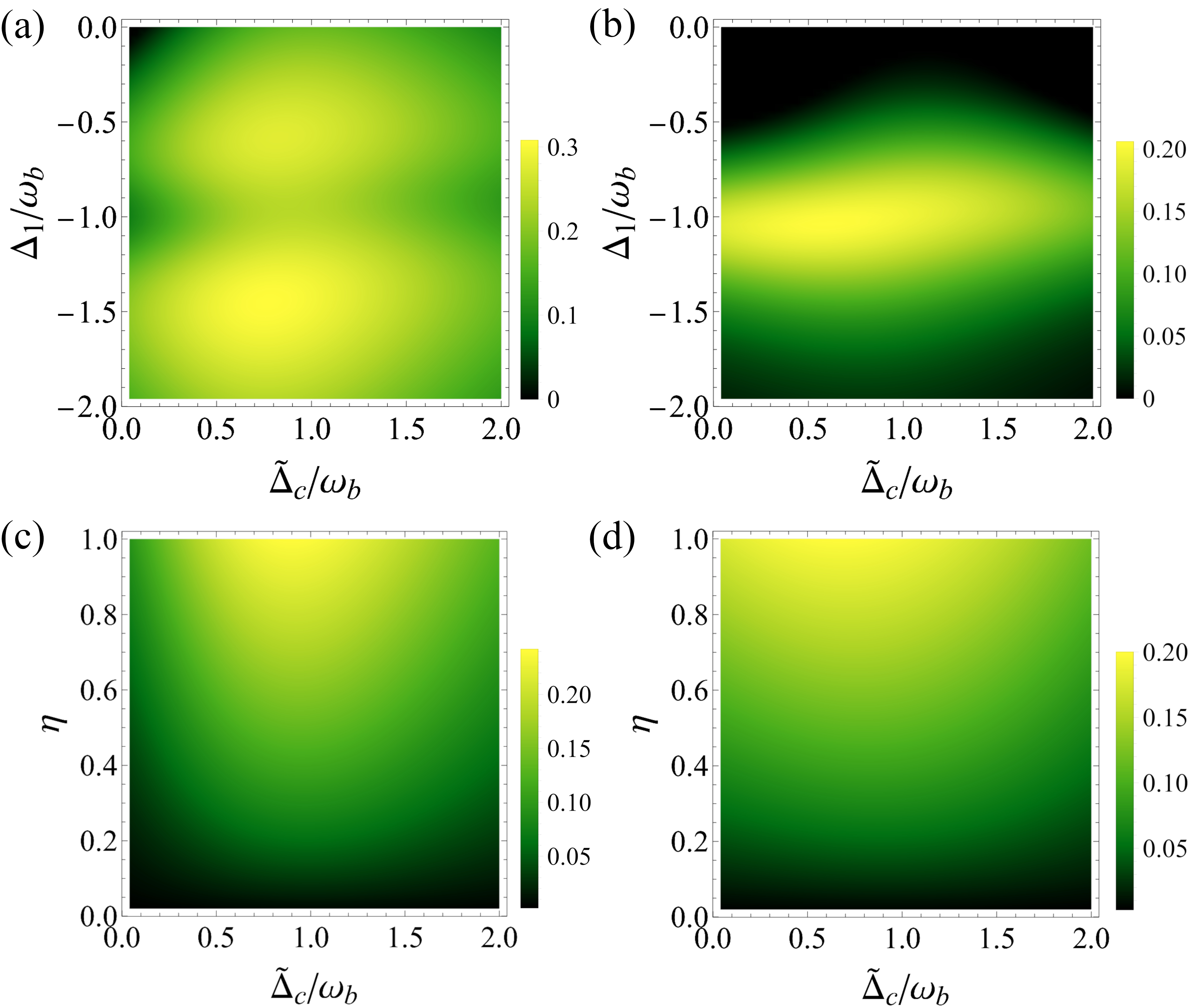}
	\caption{Density plot of steady-state (a) $E_{a_1b}$ and (b) $E_{mb}$ versus the detunings $\tilde{\Delta}_c$ and $\Delta_1$; (c) $E_{a_1b}$ and (d) $E_{mb}$ versus the detuning $\tilde{\Delta}_c$ and the coupling efficiency $\eta$. Note that we take $\eta=1$ in ($\mathrm{a}$) and ($\mathrm{b}$), and $\Delta_1/\omega_b=-1$ in  ($\mathrm{c}$) and ($\mathrm{d}$). See main text for other plotting parameters.}
	\label{fig2}
\end{figure}

Here, based on this hybrid system, we provide an accessible scheme to realize quantum entanglement of two distant mechanical nodes with resonant frequencies in completely different frequency ranges. As depicted in Fig.\ \ref{fig1}(b), we adopt the mechanism to redistribute the quantum entanglement generated in the upstream system including low-frequency mechanical mode to the downstream system with a high-frequency mechanical resonance in the network. In this hybrid model, the mechanical mode supported by the upstream optomechanical system owns the megahertz resonant frequency, and takes high thermal population even at cryogenic temperatures, which inhibits the generation of quantum entanglement. To overcome this, following the pioneering works \cite{Vitali2007,Genes2008,Li2018}, we apply a {\it sufficiently strong red-detuned} optical drive field to cool the mechanical mode and simultaneously entangle it with the cavity mode $c$, shown by the inset (1) in Fig.\ \ref{fig1}(a). Specifically, such a strong drive enables to yield an enhanced optomechanical coupling $G_c$, which can not only strengthen the optomechanical anti-Stokes scattering and activate the beam-splitter interaction $H_{aS}\propto bc^\dagger + b^\dagger c$ to greatly reduce the mechanical thermal effect, but also highlight the role played by the Stokes scattering to stimulate parametric-down conversion interaction $H_{S}\propto bc + b^\dagger c^\dagger$, owing to that the strong optomechanical coupling $G_c$ breaks the weak-coupling condition $G_c \ll \omega_b$ to take the rotating-wave approximation (RWA) in the general red-detuned driving case \cite{Vitali2007,Aspelmeyer2014,Kippenberg2008,Aspelmeyer2010}. Therefore, the optical cavity output field transmitted to the downstream optomechanical system can be filtered by the corresponding optical cavity with WGM $a_1$ resonant to the optical down-conversion sideband component which exhibits the most significant entanglement with the low-frequency mechanical mode $b$. The established quantum correlation can be further distributed to the downstream optomechanical system embodying the high-frequency mechanical node. To achieve this, we choose to utilize the approach manifested in inset (2) of Fig.\ \ref{fig1}(a). Correspondingly, the optical WGM $a_2$ is pumped by a resonant drive ($\Delta_2=0$ and thus $\Delta_m=\omega_m+\omega_0-\omega_l=\omega_1-\omega_l=\Delta_1$), which can effectively activate the optomechanical state-transfer interaction between the optical WGM $a_1$ and the gigahertz mechanical mode $m$. As a result, the quantum entanglement carried by the output field of the left optomechanical subsystem in Fig.\ \ref{fig1}(a) can be further distributed to the high-frequency mechanical node, hence realizing the remote macroscopic entanglement between two types of distinct mechanical systems with a far-detuned resonant frequency mismatch.

To quantify bipartite entanglement in the hybrid system, we consider the logarithmic negativity $E_N$ \cite{Vitali2007,Adesso2004} as the quantum measure, which can be obtained from the CM of any two-mode subsystem. In Fig.\ \ref{fig2}(a) (Fig.\ \ref{fig2}(b)), we show the bipartite entanglement $E_{a_1b}$ ($E_{mb}$) versus the detunings $\tilde{\Delta}_c$ and $\Delta_1$, by employing experimentally feasible parameters \cite{Aspelmeyer2014,Shen2022,Vanner2021,Vanner2021B}: $\omega_b/2\pi=10\ {\mathrm{MHz}}$, $\omega_m/2\pi=8.2\ {\mathrm{GHz}}$, $\lambda_c=\lambda_{a_1}\approx \lambda_{a_2}\approx 1550\ \mathrm{nm}$, $\kappa_c=\kappa_a=2.0\ \omega_b$, $\gamma_b=10^{-4}\omega_b$, $\gamma_m=0.5\omega_b$, $g_c^{}/2\pi=10^2\ \mathrm{Hz}$, $g/2\pi=20\ \mathrm{Hz}$, $\eta=1$, $T=10\ \mathrm{mK}$. Besides, in plotting Figs.\ \ref{fig1}(a) and \ref{fig1}(b), we mainly study the quantum entanglement established under fixed coupling strengths: $|G_c|/2\pi=|G_2|/2\pi= 3.0\ \mathrm{MHz}$, which can be achieved via the specific optical drivings $\epsilon$ and $\epsilon_2$ at different detunings $\tilde{\Delta}_c$ and $\Delta_1$ (see Eq. \eqref{stdystate} about the steady-state excitations). Meanwhile, the coupling strengths $G_1$ and $G_m$ can also be correpondingly determined. Clearly, exploiting the strong red-detuned pump $\tilde{\Delta}_c\approx \omega_b$, the steady-state entanglement $E_{a_1b}$ and $E_{mb}$ significantly emerge near the down-conversion sideband ($\omega_1\approx\omega_l-\omega_b$), which confirms the entanglement mechanism we discussed above. Another typical sign in these two subfigures is the complementary entanglement distribution between them. It reflects that the entanglement behaves like a finite quantum resource. Figs. \ref{fig2}(c) and \ref{fig2}(d) respectively show $E_{a_1b}$ and $E_{mb}$ along with the variation of coupling efficiency $\eta$. The results validates our protocol in remote preparation of mechanical entanglement. Here, we note that all results presented in this section are obtained in the steady state and guaranteed by the negative eigenvalues (real
parts) of the drift matrix $A$ and its submatrixes of two optomechanical systems.
 
 \begin{figure}[h]
 	\centering
 	\includegraphics[width=1\linewidth]{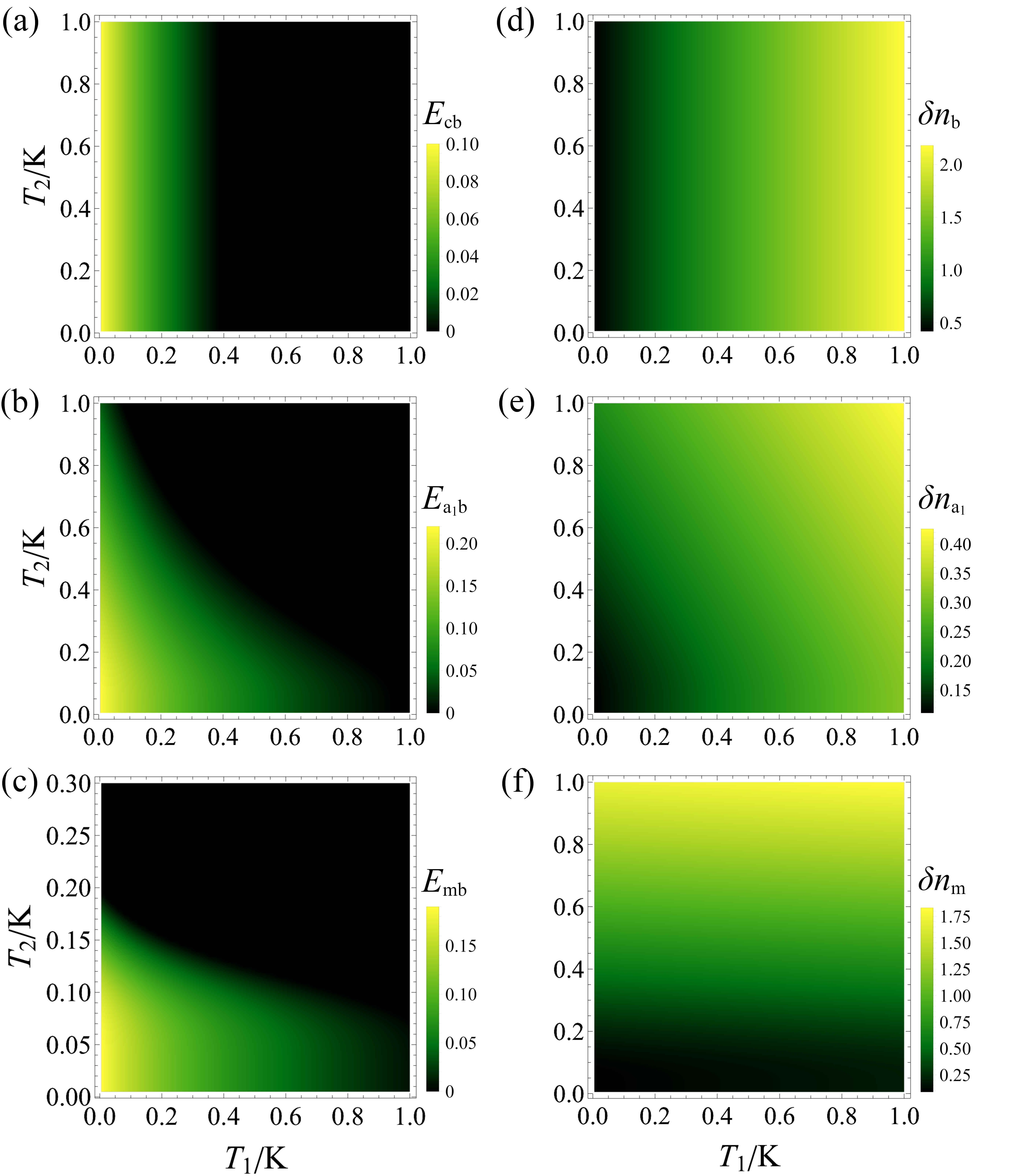}
 	\caption{Density plot of bipartite entanglement ($\mathrm{a}$) $E_{cb}$, ($\mathrm{b}$) $E_{a_1b}$ and ($\mathrm{c}$) $E_{mb}$; and quantum excitations ($\mathrm{d}$) $\delta n_{b}$, ($\mathrm{e}$) $\delta n_{a_1}$ and ($\mathrm{f}$) $\delta n_{m}$ versus the environmental temperatures $T_1$ and $T_2$. Note that here we take $\eta=0.9$, $\tilde{\Delta}_c/\omega_b=0.75$, and $\Delta_1/\omega_b=-1$. The other parameters are the same as those used in Fig. \ref{fig2}.}
 	\label{fig3}
 \end{figure}

We also discuss the influence of thermal effect to our entanglement scheme. In Figs.\ \ref{fig3}(a)-\ref{fig3}(c), we plot the steady-state entanglement $E_{cb}$, $E_{a_1b}$ and $E_{mb}$ established via this scheme when the optomechanical systems are respectively placed in the environment with temperature $T_1$ and $T_2$. Obviously, we find that $E_{cb}$ remains robust against the change of $T_2$ attributed to the unidirectional coupling between the cavity modes $c$ and $a_1$. In contrast, $E_{a_1b}$ and $E_{mb}$ are more sensitive to the variation of $T_2$ (even surpass the influence of $T_1$ in Fig. \ref{fig3}(c)). This might be a bit of  counter-intuitive because that the megahertz mechanical mode usually experiences more thermal noise than the high-frequency one in the gigahertz range, thus the quantum entanglement should exists more robust to the varying of $T_2$. Actually, this can be explained when we analyze the intrinsic quantum excitaions $\delta n_i=(\langle \delta X_i^2\rangle +\langle \delta Y_i^2\rangle-1)/2$ ($i=b,a_1$ and $m$) embodied in the quantum quadratures variances of each mode in the steady state (see Figs.\ \ref{fig3}(d)-\ref{fig3}(f)). As expressed by Fig.\ \ref{fig1}(b), the quantum dynamics of $a_1$ is not only affected by the output field of cavity mode $c$ which brings the nonlocal quantum entanglement and thus can be equivalently recognized as an external quantum reservoir, but also influenced by the thermal noise from the mechanical mode $m$. Once the temperature $T_2$ increases, the thermal effect of $m$ becomes more remarkable. However, not as the upstream optomechanical system with effective cooling mechanism activated by the stong red-detuned pump, in the downstream optomechanical system the cooling mechanism is absent, and the mechanical thermal effect tends to dissipate into the physical systems (optical WGMs) with larger dissipation rates, as in Fig.\ \ref{fig3}(e), which gives rises to entanglement dependence on $T_2$. For high-frequency mechanical node $m$, the thermal noise arising from increasing $T_2$ is more significant than that from $T_1$ (see Fig. \ref{fig3}(f)). As a result, the mechanical entanglement $E_{mb}$ is more insensitive to the variation of $T_1$ than $T_2$.

%between the low-frequency mechanical mode $b$ and the WGM $a_1$ (high-frequency mechanical mode $m$)
%Such two couplings intuitively correspond to the specific optical driving strengths $E$ and $E_2$ (see Eqs. \eqref{stdystate} for classical averages), which can further determine $G_1$ and $G_m$ in the correspoding case.

\section{Remote entanglement distribution from high- to low-frequency mechanical nodes}
In the last section, we have introduced a protocol to achieve the macroscopic entanglement of two mechanical nodes by transmitting the optical field which brings the quantum entanglement with the megahertz mechanical mode to the remote optomechanical system supporting gigahertz mechanical mode. On the other hand, how to achieve the quantum entanglement redistribution from the high- to low-frequency mechanical nodes is also worthy of study. Specifically, the scheme we proposed in last section is not subject to this case, owing to that: for the protocol we discussed in Fig.\ \ref{fig1}(a), because of the absence of cooling mechanism, the thermal effect exhibited by mechanical node in the downstream optomechanical system needs to be strictly restricted. However, under the condition that the mechanical resonant frequency lies in the megahertz range, the mechanical thermal effect can act to prevent the establishment of the distant mechanical entanglement even in the cryogenic temperature.
%reasons upon two aspects: {\it i}) For the protocol we discussed in Fig. \ref{fig1}(a), because of the absence of cooling mechanism, the thermal effect exhibited by mechanical mode in the right optomechanical subsystem needs to be strictly restricted. However, under the circumstance that the mechanical resonant frequency lies in the megahertz range, the mechanical thermal effect will hinder the establishmenet of quantum entanglement. {\it ii}) From perspective of the experimental feasibility, the successful realization of distant mechanical entanglement requires a strong coupling $G_{cM}$ which is comparable to the mechanical resoant frequency $\omega_b$ (in gigahertz range) to break RWA and activate the down-conversion component within the output field ($\kappa_c \gtrsim \omega_b$). To reach this, we have to consider a continuous pump of sevral hundred milliwatts even for the optomechanical crystal which possesses the most significant single-photon coupling $g_c^{}$ unitl now.

\begin{figure}[t]
	\centering
	\includegraphics[width=1.02\linewidth]{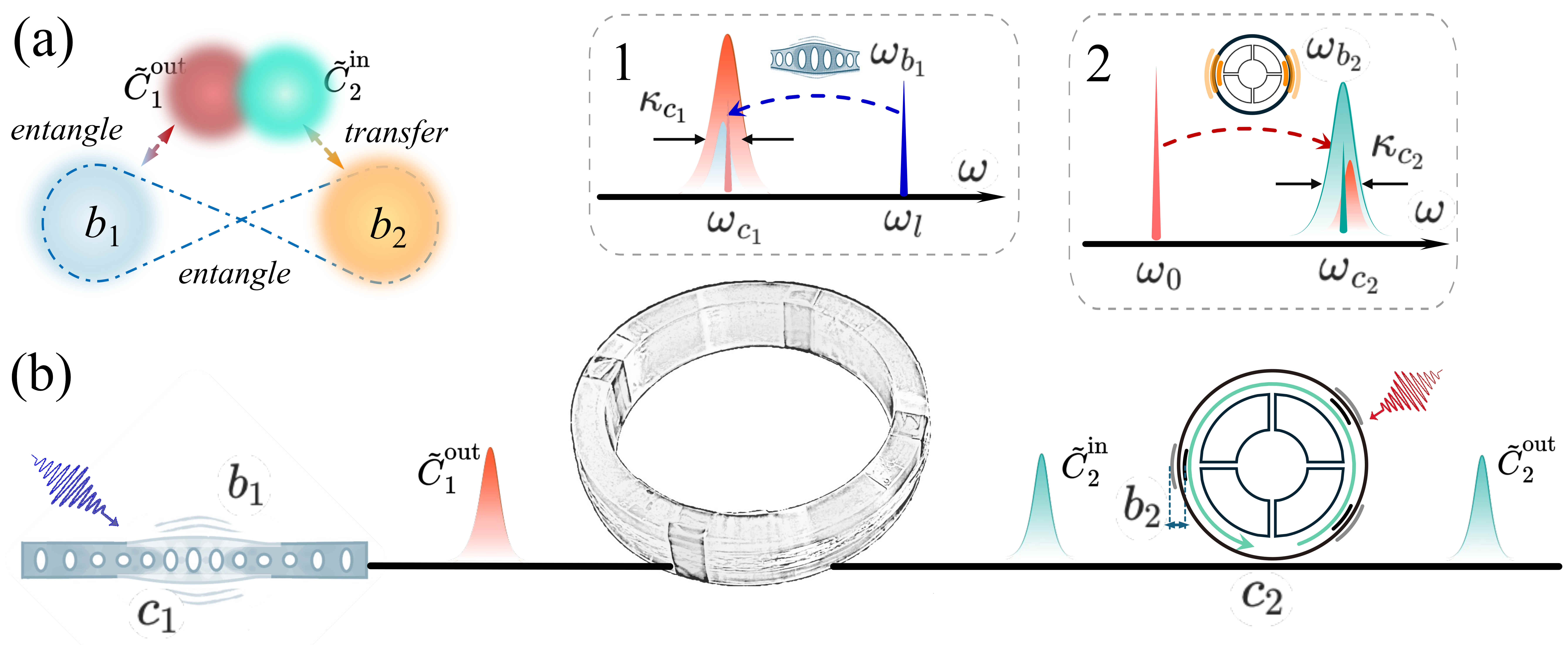}
	\caption{Schematic diagrams of the ($\mathrm{a}$) physical mechanism and ($\mathrm{b}$) hybrid system to realize the remote gigahertz-to-megahertz mechanical entanglement distribution. Inset 1: Entanglement preparation process. A blue-detuned optical pulse is used to activate the quantum entanglement between the optical output temporal mode ($\tilde{C}_1^{\mathrm{out}}$) and the gigahertz mechanical node ($b_1$). Inset 2: Entanglement distribution process. After the long-distance propagation in optical waveguide, the quantum entanglement can be further distributed to megahertz mechanical node ($b_2$) via the beam-splitter interaction activated by the red-detuned pulse.}
	\label{fig4}
\end{figure}
\subsection{Theoretical Proposal}

We proceed a new approach to achieve entanglement distribution from the high-frequency mechanical node to the remote low-frequency one, by exploiting the fast optical pulses with optomechanical system \cite{Riedinger2016,Riedinger2018}. In this scheme, we consider a hybrid model which is comprised of two optomechanical systems linked by a long-distance optical waveguide. As shown in Fig.\ \ref{fig4}, the upstream system, e.g. the optomechanical crystal \cite{Aspelmeyer2014,Riedinger2016,Riedinger2018,Hong2017,Fiaschi2021,Marinkovic2018,Chan2011}, supports the mechanical mode with resonant frequency in the gigahertz range and holds nonliear dispersive coupling. Its Hamiltonian consists of three terms $H'=H_0'+H_{\mathrm{int}}'+H_{\mathrm{dri+}}'$, including  free Hamiltonian $H_0'=\hbar\omega_{c_1} c_1^\dagger c_1+\hbar \omega_{b_1} b_1^\dagger b_1$, interaction $H_{\mathrm{int}}'=-\hbar g_1^{} c_1^\dagger c_1(b_1 + b_1^\dagger)$, and driving $H_{\mathrm{dri}+}'=i\hbar \epsilon'(c_1^\dagger e^{-i\omega_l t}-c_1 e^{i\omega_l t})$. This presents a ideal platform to prepare the optomechanical entangled state using nonlinear coupling. Therefore, we choose to enhance the optomechanical Stokes scattering by sending a blue-detuned optical pulse  ($\omega_l=\omega_{c_1}+\omega_{b_1}$) to the upstream optomechanical system. In resolved sideband regime, this can activate a parametric-down conversion interaction $H_b=-\hbar G_b(b_1 c_1 + b_1^\dagger c_1^\dagger)$ between the cavity photons and the gigahertz mechanical phonons, which dominates the dynamics of the optomechanical system during the pulse operation and leads to following QLEs in the interaction picture:
\begin{equation}
	\begin{split}
		\dot{c}_1=&\ iG_b b_1^\dagger -\frac{\kappa_{c_1}}{2}c_1 +\sqrt{\kappa_c} c_{1}^{\mathrm{in}},\\
		\dot{b}_1\approx &\ iG_b c_1^\dagger.
	\end{split}
\end{equation}
Specifically, we considered a {\it weak} flattop pulse with duration much shorter than the mechanical lifetime, thus the dissipation of mechanical mode can be regarded to be negligibly small and we can safely neglect it to simplify the model. Given by weak coupling $G_b\ll \kappa_{c_1}$, the dynamics of the cavity mode can be adiabatically eliminated to obtain $c_1\approx 2/\kappa_{c_1}(iG_b b_1^\dagger+\sqrt{\kappa_{c_1}} c_{1}^\mathrm{in})$. Accounting for the cavity input-output relation: $c_{1}^\mathrm{out}=\sqrt{\kappa_{c_1}}c_1-c_{1}^\mathrm{in}$ \cite{Gardiner1985}, we can obtain 
\begin{equation}
	\label{left}
	\begin{split}
		c_{1}^\mathrm{out}=&\ i\sqrt{2\mathcal{G}_b}b_1^\dagger+ c_{1}^\mathrm{in},\\
		\dot{b}_1= &\ -\mathcal{G}_b b_1+i\sqrt{2\mathcal{G}_b } c_{1}^\mathrm{out},
	\end{split}
\end{equation}
where $\mathcal{G}_b$=$2G_b^2/\kappa_{c_1}$. Following the pioneering studies in cavity optomechanics \cite{Hofer2011,Galland2014} and related cavity magnonics \cite{Li2021,Lu2025}, we can describe the physical process of entanglement preparation via defining a set of normalized temporal modes for the cavity mode driven by a pulse with duration $\tau_b$, which gives that 
\begin{subequations}
		\label{temporal}
		\begin{align}
			\tilde{C}_1^\mathrm{in}(\tau_b)=&\sqrt{\frac{2\mathcal{G}_b}{1-e^{-2\mathcal{G}_b\tau_b}}}\int_{0}^{\tau_b} e^{-\mathcal{G}_b s}c_{1}^\mathrm{in}(s)\ ds, \\
			\label{temporalout}
			\tilde{C}_1^\mathrm{out}(\tau_b)=&\sqrt{\frac{2\mathcal{G}_b}{e^{2\mathcal{G}_b\tau_b}-1}}\int_{0}^{\tau_b} e^{\mathcal{G}_b s}c_{1}^\mathrm{out} (s)\ ds.
		\end{align}
\end{subequations}
Equipped with the necessary tool \eqref{temporal}, we can integrate $\eqref{left}$ to get:
\begin{equation}
	\label{entangle}
	\begin{split}
		\tilde{C}_1^\mathrm{out}(\tau_b)=&\ i\sqrt{e^{2\mathcal{G}_b\tau_b}-1}b_1^\dagger(0)+e^{\mathcal{G}_b\tau_b}\tilde{C}_1^\mathrm{in}(\tau_b),\\
		b_1(\tau_b)=&\ e^{\mathcal{G}_b\tau_b}b_1(0)+i\sqrt{e^{2\mathcal{G}_b\tau_b}-1}\tilde{C}_1^{\mathrm{in}\dagger}(\tau_b),
	\end{split}
\end{equation}
and extract the corresponding propagator \cite{Galland2014}
\begin{equation}
	U_b(\tau_b)=e^{i\mathrm{tanh}r\tilde{C}_1^{\mathrm{in}\dagger} b_1^\dagger } \mathrm{cosh}r^{-1-\tilde{C}_1^{\mathrm{in}\dagger}\tilde{C}_1^{\mathrm{in}}-b_1^\dagger b_1^{} }e^{-i\mathrm{tanh}r \tilde{C}_1^{\mathrm{in}} b_1},
\end{equation}
which allows the effective transformations from $b(0)$ and $\tilde{C}_1^\mathrm{in}(0)$ to $b(\tau_b)$=$U_b^\dagger(\tau_b)b(0)U_b(\tau_b)$ and $\tilde{C}_1^\mathrm{out}(\tau_b)$=$U_b^\dagger(\tau_b)\tilde{C}_1^\mathrm{in}(\tau_b)$ $ U_b(\tau_b)$. Here, we simplified the expression by $\mathrm{cosh}r\ $=$\ e^{\mathcal{G}_b\tau_b}$, $\mathrm{tanh}r\ $=$\sqrt{1-e^{-2\mathcal{G}_b\tau_b}}$, both of which includes the squeezed parameter $r$ characterizing the parametric-down conversion-related squeezing strength.

Considering $\rho_{b_1}(0)$ to be the initial state of low-frequency mechanical mode, the state of the upstream optomechanical system, at end of the blue-detuned optical pulse, can be represented by $\rho_{\tilde{C}_1 b_1}(\tau)$=$U_b(\tau)\rho_{\tilde{C}_1b_1}(0)U_b^\dagger(\tau)$=$U_b(\tau)[|0\rangle _{\tilde{C}_1} \langle0|\otimes \rho_{b_1}(0)]U_b^\dagger(\tau)$. To achieve the quantum entanglement, we assume that the gigahertz mechanical mode is precooled to its ground state, i.e. the mechanical initial state $\rho_{b_1}(0)=|0\rangle _{b_1} \langle0|$, which has been successfully demonstrated in the experimental platform based on the optomechanical crystal \cite{Chan2011,Painter2015}. Therefore, after a pulse period of $\tau_b$, the quantum state 
\begin{equation}
	\begin{split}
		\raisetag{17pt}
		&\rho_{\tilde{C}_1 b_1}(\tau_b)=\\
		&\ \ \ \ \  (\mathrm{cosh}r)^{-2}\sum_{n=0}^{\infty}\sum_{n'=0}^{\infty}(i\mathrm{tanh}r)^n|n,n\rangle_{\tilde{C}_1^{\mathrm{out}}b_1}\langle n',n'|(-i\mathrm{tanh}r)^{n'},
	\end{split}
\end{equation}
is a two-mode squeezed vacuum state referring to the high-frequency mechanical mode and the optical output temporal mode. Such state, intrinsically embodying the nature of quantum entanglement, can be characterized by the quantum measure of logarithmic negativity $E_N=2r$.			

As illustrated by the sketch in Fig.\ \ref{fig4}, with the mediation of the optical waveguide, the output temporal pulse can propagate towards the downstream optomechanical system, which holds the low-frequency mechanical mode in the megahertz range. In order to realize the entanglement distribution onto the low-frequency mechanical node, this system is driven by a red-detuned optical pulse to excitate the beam-splitter interaction $H_r=-\hbar G_r(b_1 c_1^\dagger + b_1^\dagger c_1)$ . In this context, we mainly consider the optomechanical configuration of microresonators \cite{Aspelmeyer2014} and assume that the duration of used fast optical pulse is much smaller than the coherent lifetime of the mechanical mode $\tau_r$$\ll$$1/\gamma_b$. Therefore, we can safely exclude the dissipation-related terms in the QLEs and simplify the model by adiabatically eliminated the optical cavity mode $c_2$ to get $c_2\approx 2/\kappa_{c_2}(iG_r b_2+\sqrt{\kappa_{c_2}} c_{2}^\mathrm{in})$. Similar to  \eqref{left} in the entanglement generation process, we can obtain  
\begin{equation}
	\label{right}
	\begin{split}
		c_{2}^\mathrm{out}=&\ i\sqrt{2\mathcal{G}_r}b_2+ c_{2}^\mathrm{in},\\
		\dot{b}_2= &\ \mathcal{G}_r b_2+i\sqrt{2\mathcal{G}_r } c_{2}^\mathrm{out},
	\end{split}
\end{equation}
by combining the cavity input-output relation $c_{2}^\mathrm{out}=\sqrt{\kappa_{c_2}}c_2-c_{2}^\mathrm{in}$, with $\mathcal{G}_r=2G_r^2/\kappa_{c_2}$. Following the integration process to get \eqref{entangle}, this achieve a beam-splitter type $2\times 2$ unitary transformation
\begin{equation}
	\label{bstransformation}
	\begin{split}
		\tilde{C}_2^\mathrm{out}(\tau_r)=&\ i\sqrt{1-e^{-2\mathcal{G}_r\tau_r}}b_2(0)+e^{-\mathcal{G}_r\tau_r}\tilde{C}_2^\mathrm{in}(\tau_r),\\
		b_2(\tau_r)=&\ e^{-\mathcal{G}_r\tau_r}b_2(0)+i\sqrt{1-e^{2\mathcal{G}_r\tau_r}}\tilde{C}_2^{\mathrm{in}}(\tau_r),
	\end{split}
\end{equation}
by defining of a set of normalized temporal modes
\begin{subequations}
	\label{C2tem}
	\begin{align}
		\label{C2temin}
		\tilde{C}_2^\mathrm{in}(\tau_r)=&\sqrt{\frac{2\mathcal{G}_r}{e^{2\mathcal{G}_r\tau_r}-1}}\int_{0}^{\tau_r} e^{\mathcal{G}_r s'}c_{2}^\mathrm{in}(s')\ ds', \\
		\tilde{C}_2^\mathrm{out}(\tau_r)=&\sqrt{\frac{2\mathcal{G}_r}{1-e^{-2\mathcal{G}_r\tau_r}}}\int_{0}^{\tau_r} e^{-\mathcal{G}_r s'}c_{2}^\mathrm{out} (s')\ ds'.
	\end{align}
\end{subequations}
Briefly introduced the entanglement proposal sketched in Fig.\ \ref{fig4}(a), afterwards we can specifically discuss the quantum state in subsequent process of entanglement distribution.

\subsection{Entanglement Redistribution}
\begin{figure}[h]
	\centering
	\includegraphics[width=0.5\linewidth]{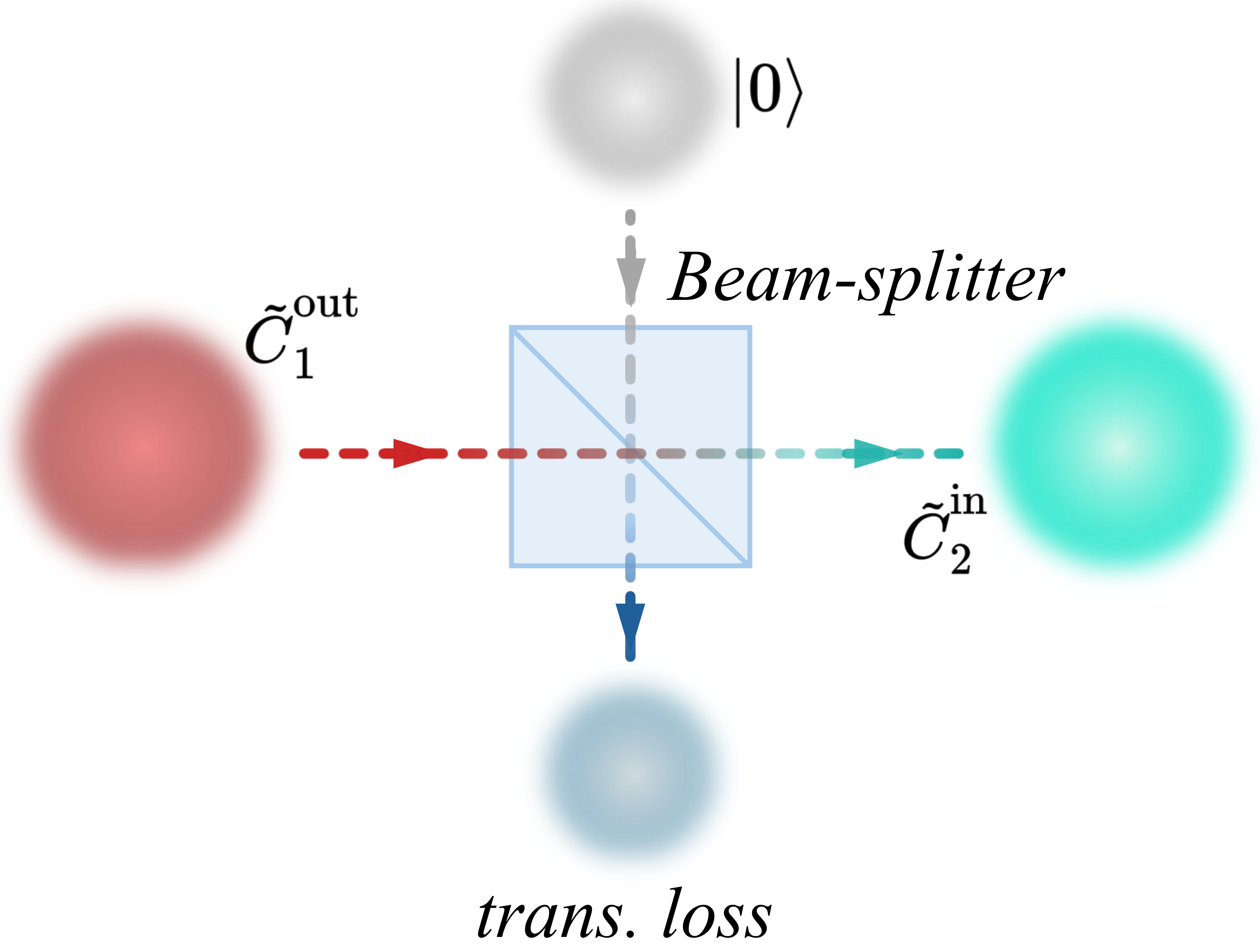}
	\caption{Sketch of the transmission loss model.}
	\label{figbs}
\end{figure}

%\begin{comment}
For building a long-distance quantum network which connects two distinct mechanical nodes via the optical mediation, the transmission loss is practically unavoidable. The optical loss significantly increases along with the transmission distance, therefore it can be theoretically described by the linear beam-splitter model \cite{Leonhardt1997,Li2010,Li2021}, where $R$ and $T$ respectively represent the reflectance and transmittance of the beam-splitter and satisfy $R+T=1$. Naturally, this transformation provides a clear relation about the output temporal mode $\tilde{C}_1^\mathrm{out}$ and the input temporal mode $\tilde{C}_2^\mathrm{in}$ (see Fig.\  \ref{figbs}). With the bridge of it, we can turn into the quantum state after a transmitting time of $\Delta\tau$:
	\begin{equation}
		\begin{split}
			\raisetag{15pt}
			&\rho_{\tilde{C}_2 b_1}(\tau_b+\Delta\tau)= \\
			&\ \ \ \ (\mathrm{cosh}r )^{-2}\sum_{n=0}^\infty\sum_{n'=0}^{\infty} (i\mathrm{tanh}r)^n(-i\mathrm{tanh}r)^{n'}|n\rangle_{b_1}\langle n'|\ \otimes\ \\
			&\ \ \ \ \sum_{t=0}^{\mathrm{min}(n,n') }\sqrt{\frac{n!n'!}{(t!)^2(n-t)!(n'-t)!} } R^tT^{\frac{n+n'}{2}-t }|n-t\rangle_{\tilde{C}_2^{\mathrm{in}}}\langle n'-t|.
		\end{split}
	\end{equation}
%\end{comment}

Expressed by \eqref{bstransformation}, the pulse-assisted state transfer process introduces the output temporal mode $\tilde{C}_2^{\mathrm{out}}$ and low-frequency mechanical mode $b_2$ which is assumed to be precooled to its ground state \cite{Park2009,Schliesser2008,Schliesser2009,Verhagen2012}. The initial state, before activating the optomechanical beam-splitter interaction in the downstream optomechanical system, can arrive at
\begin{equation}
	\rho_{\mathrm{BS} }^{}(0)=\rho_{\tilde{C}_2 b_1}(\tau_b+\Delta\tau)\otimes|0,0\rangle_{\tilde{C}_2^{\mathrm{out}}b_2}\langle 0,0|.
\end{equation}

As a result, the transformation \eqref{bstransformation} enables us to obtain the quantum state after a red-detuned pulse of $\tau_r$
\begin{equation}
	\begin{split}
		\rho_{\mathrm{BS}}(\tau_r)=&\ (\mathrm{cosh}r )^{-2}\sum_{n=0}^{\infty}\sum_{n'=0}^{\infty}(i\mathrm{tanh}r )^n(-i\mathrm{tanh}r )^{n'} |n\rangle_{b_1}\langle n|\\
		&\otimes\sum_{t=0}^{\mathrm{min}(n,n') }\sqrt{\frac{n!n'!}{(t!)^2(n-t)!(n'-t)!} }R^tT^{\frac{n+n'}{2}-t }\\
		&\sum_{q=0}^{n-t }\sum_{p=0}^{n'-t }\sqrt{\frac{(n-t)!(n'-t)!}{q!p!(n-t-q)!(n'-t-p)!} }\\
		&i^q(-i)^p W^{\frac{p+q}{2}}(1-W)^{\frac{n+n'-p-q}{2}-t }\ |q\rangle_{b_2}\langle p|\\
		&\otimes|n-t-q\rangle_{\tilde{C}_2^{\mathrm{out}}}\langle n'-t-p|\otimes|0\rangle_{\tilde{C}_2^{\mathrm{in}}}\langle 0|,
	\end{split}
\end{equation}
with the optomechanical state transfer efficiency $W=1-e^{-2\mathcal{G}_r\tau_r}$.
By tracing out unrelated modes, we can extract the density matrix which characterizes the subsystem of two mechanical nodes in this network:
\begin{widetext}
	\begin{equation}
		\label{mechanicalsub}
		\begin{split}
			\rho_{\mathrm{sub}}&= \rho_{b_1b_2}^{}=\mathrm{Tr}_{\tilde{C}_2}\left[\rho_{\mathrm{BS}}(\tau_r)\right]			\\
			&=(\mathrm{cosh}r )^{-2}\sum_{n=0}^{\infty}\sum_{n'=0}^{\infty}(i\mathrm{tanh}r )^n(-i\mathrm{tanh}r )^{n'} |n\rangle_{b_1}\langle n|\otimes\sum_{t=0}^{\mathrm{min}(n,n') }\sqrt{\frac{n!n'!}{(t!)^2(n-t)!(n'-t)!} }R^tT^{\frac{n+n'}{2}-t }\\
			&\sum_{l=0}^{\mathrm{min}(n-t,n'-t) }\sqrt{\frac{(n-t)!(n'-t)!}{(l!)^2(n-t-l)!(n'-t-l)!} }\ i^{n-t-l}(-i)^{n'-t-l} W^{\frac{n+n'}{2}-t-l}(1-W)^l|n-t-l\rangle_{b_2}\langle n'-t-l|,
		\end{split}
	\end{equation}
\end{widetext}
%from which we can confirm the macroscopic entanglement between two mechanical nodes with significantly different resonant frequencies. 
from which we can successfully derive the CM corresponding to this two-mode state. Using the CM definition $V_{ij}=\frac{1}{2}\langle \xi_i \xi_j+\xi_j\xi_i\rangle -\langle \xi_i\rangle \langle \xi_j\rangle =\frac{1}{2}[\mathrm{Tr}(\rho_{\mathrm{sub}}\xi_i\xi_j)+\mathrm{Tr}(\rho_{\mathrm{sub}}\xi_j\xi_i)]-\mathrm{Tr}(\rho_{\mathrm{sub}}\xi_i)\times\mathrm{Tr}(\rho_{\mathrm{sub}}\xi_j)$, where $i,j\in\{1,2,3,4\}$ and $\xi=\{\delta X_{b_1}$, $\delta Y_{b_1}$, $\delta X_{b_2}$, $\delta Y_{b_2}\}^{\mathrm{T}}$, the calculated CM reads
\begin{equation}
	\begin{split}
		\raisetag{-35pt}
		&V_{\mathrm{sub}}=\\
		&\ \ \begin{pmatrix}
			\frac{1}{2}\mathrm{cosh}2r  & 0 & -\frac{\sqrt{WT}}{2}\mathrm{sinh}2r  & 0\\
			0 & \frac{1}{2}\mathrm{cosh}2r & 0 & \frac{\sqrt{WT}}{2}\mathrm{sinh}2r\\
			-\frac{\sqrt{WT}}{2}\mathrm{sinh}2r & 0 & WT\mathrm{sinh}^2r +\frac{1}{2} & 0\\
			0 & \frac{\sqrt{WT}}{2}\mathrm{sinh}2r & 0 & WT\mathrm{sinh}^2r+\frac{1}{2}
		\end{pmatrix}.
	\end{split}
\end{equation}

\begin{figure}[h]
	\centering
	\vspace{8pt}
	\hspace{-40pt}
	\includegraphics[width=0.55\linewidth]{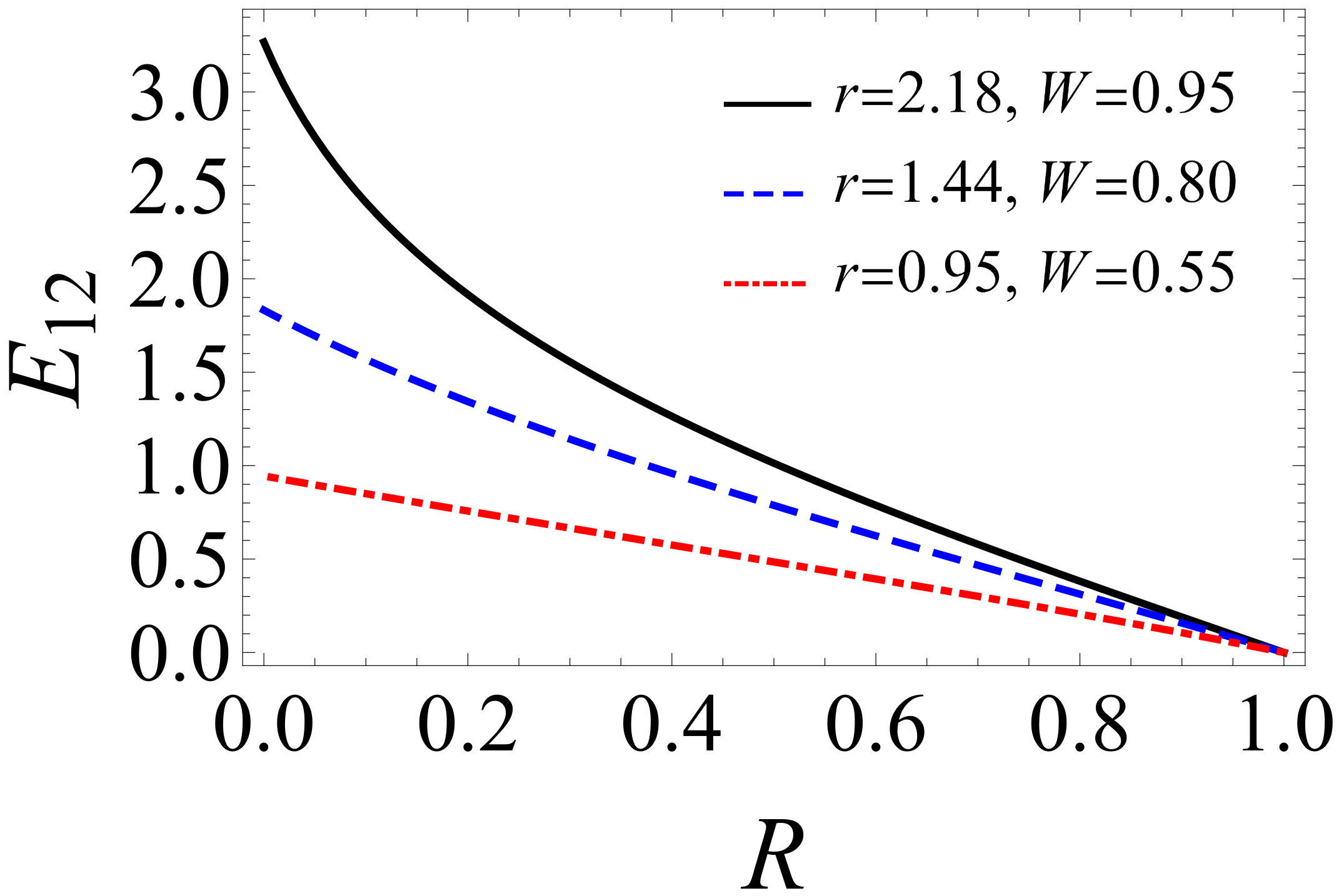}
	\caption{The macroscopic entanglement $E_{12}$ versus the reflectivity (optical loss) $R$ under different conditions, with two-mode squeezing and state conversion parameters: $r=2.18$ and $W=0.95$ (solid), $r=1.44$ and $W=80$ (dashed), and $r=0.95$ and $W=0.55$ (dot-dashed).}
	\label{fig6}
\end{figure}

Following section \ref{sec2}, we adopt the quantum measure logarithmic negativity $E_{12}$ to characterize the macroscopic entanglement. In Fig.\ \ref{fig6}, we  show the varying of entanglement $E_{12}$ versus the optical loss $R$ under different two-mode squeezing and state conversion parameters, respectively, which reveals the restriction of transmission distance (considering fiber loss 0.2 $ \mathrm{dB}/\mathrm{km}$ at the telecom wavelength \cite{Li2021}) on achieved  entanglement between two mechanical nodes.
% Moreover, it should be remarked that, to maintain the shape consistency of the optical temporal modes $\tilde{C}_1^{\mathrm{out}}$ and $\tilde{C}_2^{\mathrm{in}}$ (see Eqs.\ \eqref{temporalout} and \eqref{C2temin}), the parameters $\mathcal{G}_r$ and  $\tau_r$ determined by the red-detuned pulse applied in the entanglement distribution process are kept same with $\mathcal{G}_b$ and $\tau_b$ used in the entanglement preparation process, i.e. $\tau_r=\tau_b$, $\mathcal{G}_r=\mathcal{G}_b$, which forms the intuitive relation between the transfer efficiency and the squeezing parameter: $W$=$1-e^{-2\mathcal{G}_r \tau_r}$=$1-e^{-2\mathcal{G}_b \tau_b}$=$\mathrm{tanh}^2 r$. Therefore, the Fig.\ \ref{fig6} also gives corresponding $r$ following different transfer efficiencies.

Furthermore, we provide a brief discussion on the experimental feasibility of present entanglement scheme. In plotting Fig.\ \ref{fig6}, the optomechanical system with high-frequency mechanical mode, such as optomechanical crystal \cite{Riedinger2016}, holds telecom wavelength optical mode $\lambda_{c_1}\approx 1550\ \mathrm{nm}$, mechanical mode resonance $\omega_{b_1}/2\pi=5.3\ \mathrm{GHz}$, the optical decay rate $\kappa_{c_1}/2\pi=1.3\ \mathrm{GHz}\gg \gamma_{b_1}/2\pi$, and supports the optomechanical coupling strength up to $g_1^{}/2\pi=825\ \mathrm{kHz}$. In entanglement generation process shown in Fig.\  \ref{fig4}(b), we manage to take a blue-detuned pulse with $\tau_b=10\ \mu$s and $P'\approx 0.4\ \mathrm{\mu W}$ to realize the effective optomechanical coupling $G_1/2\pi\approx 3.93\ \mathrm{MHz}$, which gives rise to the two-mode squeezing parameter $r\approx 2.18$. On the other hand, for the entanglement redistribution process depicted by the inset (2) in Fig.\ \ref{fig4}(b), we mainly suggest the optomechanical microresonators as provided in Refs. \cite{Aspelmeyer2014,Verhagen2012}, which allow: the optical resonance near $1550\ \mathrm{nm}$, the mechanical mode $\omega_{b_2}/2\pi\approx 10^2\ \mathrm{MHz}$, optical dissipation $\kappa_{c_2}/2\pi=20\ \mathrm{MHz}\gg \gamma_{b_2}/2\pi$, and $g_2^{}/2\pi=1.7\ \mathrm{kHz}$. By employing a red-detuned optical pulse with $\tau_r=10\ \mu$s and $P''\approx 33.5\ \mathrm{\mu W}$, the optomechanical coupling $G_2$ can be greatly enhanced and proceed the optomechanical state transfer efficiency to $W=1-e^{-2\mathcal{G}_r\tau_r}\approx 0.95$. These analyzations reflect the solidity of our theoretical scheme. Moreover, the mechanical modes supported by these two kinds of optomechanical platforms have been successfully cooled to (near) their ground states \cite{Chan2011,Verhagen2012}, which also highlights feasibility of this proposal.
%With experimentally achievable parameters \cite{Riedinger2016,Verhagen2012}: $\omega_{b_1}/2\pi=5.3\ \mathrm{GHz}$, $\omega_{b_2}/2\pi=10^2\ \mathrm{MHz}$, $\lambda_{c_1}=\lambda_{c_2}=1550\ \mathrm{nm}$, $g_1^{}/2\pi=825\ \mathrm{kHz}$, $g_2^{}/2\pi=1.7\ \mathrm{kHz}$, $\kappa_{c_1}/2\pi=1.3\ \mathrm{GHz}$ and $\kappa_{c_2}/2\pi=20\ \mathrm{MHz}$,

\section{Conclusion}
In this work, we provide two schemes to study how to achieve the remote entanglement between two mechanical nodes with resonant frequencies respectively in the megahertz and gigahertz ranges, which has been much less explored before. Specifically, based on the hybrid model which includes the dispersive and triple-resonant optomechanical couplings, we first discuss a protocol to achieve the megahertz-to-gigahertz entanglement distribution. This can realized by activating the quantum entanglement in the local optomechanical system with megahertz mechanical mode and further redistribute it to the gigahertz mechanical mode via the optical photons transmission and the beam-splitter interaction activated in the distant optomechanical system. We also give a proposal to entangle two mechanical nodes over a long distance by using fast optical pulses, which suggests to redistribute the quantum entanglement generated in the optomechanical system with gigahertz mechanical mode to the remote optomechanical system with gigahertz mechanical mode. In a word, this work shows the possibility of distinct mechanical nodes in the long-distance entanglement distribution, which may broaden the further application of different cavity optomechanical platforms and mechanical systems in hybrid quantum network, quantum information processing and fundamental physics test, .

 %The various cavity optomechanical platforms and mechanical systems possess broad application in modern quantum technologies. 
\begin{acknowledgements}
We thank J. Li from Zhejiang University for helpful discussions and feedback. This work was supported by the Fundamental Research Program of Shanxi Province under Grant No. 202203021211260.
\end{acknowledgements}

%\appendix

%\section*{Appendix}

%\nocite{*}

% Produces the bibliography via BibTeX.

\end{document}